\documentclass{eas}
\usepackage{graphicx}
\begin{document}

\runningtitle{Bisector Measurements of HD~102272}
\title{The Pennsylvania-Torun Search for Planets: Bisector Measurements of HD~102272} 
\author{Grzegorz Nowak}\address{Toru\'n Centre for Astronomy, Nicolaus Copernicus University, ul. Gagarina 11, 87-100 Toru\'n, Poland}
\author{Andrzej Niedzielski$^{1,}$}\address{Department of Astronomy and Astrophysics, Pennsylvania State University, 525 Davey Laboratory, University Park, PA 16802}
\author{Aleksander Wolszczan$^{2,1}$}
%
%
\begin{abstract}
Searches for planets around massive stars are essential for developing general understanding of planet formation and evolution of the planetary systems. The main objective of the Pennsylvania - Torun Planet Search is detection of planets around G-K subgiants and giants through precision radial velocity (RV) measurements with iodine absorption cell using HET HRS spectrograph. However, the long period radial velocity variations of red giants may also have other than planetary nature (e.g. a non-radial pulsations or rotational modulation in presence of starspots). In this work we present bisector analysis of cross-correlation functions (CCF) constructed from the spectra used for radial velocity determination but cleaned from the iodine lines for the second red giant with planets from our survey HD~102272.
\end{abstract}
\maketitle
\section{Introduction}
High precision stellar radial velocity measurements are extensively used to detect the reflex motion of a star due to planetary companion. Radial velocity method, however is not sensitive only to the motion of a star around the center of mass of star-planet system. Changes in line shapes arising from stellar atmospheric motion (caused by non-radial pulsation or inhomogeneous convection and/or spots combined with rotation) or from light contamination from unseen stellar companion can mimic small radial velocity variations at the spectral resolution 50 000 - 70 000 typically utilized for planet searches. Therefore it is important (especially in case of giant stars) to investigate whether the observed radial velocity curve are caused by a shift of the spectral lines as a whole or by a change in the symmetry of the spectral lines.

\section{Observations}
The observational material used in this paper are high quality,  high-resolution optical spectra of HD~102272 observed within our survey. Observations were made with the Hobby-Eberly Telescope (HET) (Ramsey {\em et al.\/} \cite{RamseyEtAl1998}) equipped with the High Resolution Spectrograph (HRS) (Tull \cite{Tull1998}) in the queue scheduled mode (Shetrone {\em et al.\/} \cite{Shetrone2007}). The spectrograph was used in the R=60,000 resolution mode and it was fed with a 2 arcsec fiber. The spectra consisted of 46 Echelle orders recorded on the ``blue'' CCD chip (407.6 - 592 nm) and 24 orders on the ``red'' one (602 - 783.8 nm). Typical signal to noise ratio was 200-250 per resolution element. The basic data reduction and measurements were performed using standard IRAF\footnote{ IRAF is distributed by the National Optical Astronomy Observatories, which are operated by the Association of Universities for Research in Astronomy, Inc., under cooperative agreement with the National Science Foundation.} tasks and scripts.

\section{Line bisector analysis}
At present most analysis of the variations of spectral lines via line bisectors base on cross-correlation function (CCF), which represents an ``average'' spectral line of the observed star. However, in the iodine cell method of the RV determination $I_2$ lines affect stellar spectrum. We may  therefore construct the CCF only after proper removal of the iodine lines. The method for proper removal iodine lines from stellar spectra was proposed by Mart{\'{\i}}nez Fiorenzano {\em et al.\/} (\cite{MartinezFiorenzano2005}).

\subsection{Removal of {\boldmath $I_2$} lines}
In order to remove iodine lines from the stellar spectrum, all orders of the stellar spectrum are divided by the respective orders of the iodine flat field. Since echelle orders are best illuminated in the central parts only these are used for the division. We use the wavelength range were reasonably strong iodine lines appear in the HET/HRS spectra (503.6 - 590.0 nm along 17 orders). First the cross-correlation is computed between the iodine flat field  and the stellar spectrum, to determine the offset in wavelength between $I_2$ lines present in both spectra. The flat field flux is adjusted to the new wavelength scale, adding the offset previously determined, by using a Hermite spline interpolation (Hill \cite{Hill1982}). Finally the stellar spectrum is divided by iodine flat field order by order and the result is the stellar spectrum free from $I_2$ lines.

\subsection{Construction of the cross-correlation functions (CCF)}
To construct the CCF stellar spectrum is correlated with a numerical mask consisting of 1 and 0 value points, with the non-zero points corresponding to the positions of the stellar absorption lines at zero velocity. We built the numerical mask using very high SNR of HD~17092, the first star with planets from our survey (Niedzielski {\em et al.\/} \cite{Niedzielski2007}), cleaned from spectral features lying within $\pm \, 30 \; km \, s^{-1}$ from known telluric lines. The CCF of the recorded spectrum ($S(\lambda)$) is constructed by shifting the mask ($M(\lambda_{RV})$) as a function of the Doppler velocity:
\begin{equation}
CCF(RV) = \sum_{i} \int S(\lambda) M_{i}(\lambda_{RV}) d\lambda = \sum_{i} CCF_{i}(RV),
\end{equation}
where
\begin{equation}
\lambda_{RV} = \lambda \sqrt{\frac{1-\frac{RV}{c}}{1+\frac{RV}{c}}}\quad .
\end{equation}
For each order, the algorithm selects from the wavelength range and the velocity range to compute, the lines of the mask that will always be in the wavelength domain during the scan. The CCF is computed step by step for each velocity point without merging the orders. CCFs from all order are added to get the final CCF for the whole spectrum.

Altogether, about 980 spectral lines were used in the mask. In buliding the CCF no attepmpt was made to remove blended lines from the stellar spectrum. This may alter the shape of the CCF but by using many lines we are confident that the effects average out. Furthermore, as long as the same lines are used for all observations, and only variations in the shape are of interest, blended stellar lines do not affect the final result.

\subsection{Line bisector, bisector velocity span and bisector curvature}
The bisector of the CCF is the middle point of the horizontal segment connecting points on the left and right sides of the profile with the same flux value. Between some arbitrary defined minimum and maximum CCF depth values, the wings of the CCF were rebinned with a step of 0.001. For every such point bisector value was calculated. The bisector line was obtained by combining bisector points ranging from the core toward the wings of the line.

In the CCF profile we defined top, central and a low zones that represent interesting places to study the velocity given by the bisector. In choosing the span zones, it is important to avoid the wings and cores of the profile where the errors of the bisector measurements are large. For our span measurements we chose the lower zone between $5\%$ and $25\%$, central zone between $35\%$ and $55\%$, and upper zone between $65\%$ and $85\%$ in the term of the CCF depth. The changes in the spectral line bisector were quantified using the bisector velocity span ($BVS$), which is simply the velocity difference between mean bisector velocity in the upper and lower zone of the line bisector ($BVS = V_T - V_B$) and bisector curvature ($BC$) which is the difference of the velocity span of the upper half of the bisector and the lower half ($BC = (V_T - V_C) - (V_C - V_B)$). It is important to examine both $BVS$ and $BC$ because it is possible for a star to show variations in one of these quantities only. To determine errors of the $BVS$ and $BC$ we used expression for bisector velocity error given by Mart{\'{\i}}nez Fiorenzano {\em et al.\/} (\cite{MartinezFiorenzano2005}).

\section{Results}
\begin{figure}[!h]
\includegraphics[width=5.0in]{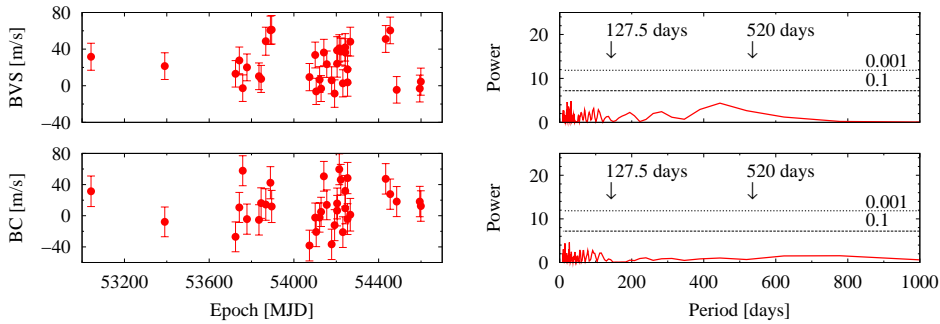}
\vskip-0.25in
\caption{BVS and BC curves and periodograms for HD~102272. The periods present in the RV data are indicated.}
\label{HD102272bis-lsp}
\end{figure}

In Figure \ref{HD102272bis-lsp} we present BVS and BC curves and their periodograms for the second red giant with planets from our survey HD~102272 (Niedzielski  {\em et al.\/} \cite{Niedzielski2008}). No significant periods whatsoever are present in these periodograms; all trial periods have extremely small significance levels. In particular, no peak is present at the RV periods, so that we conclude that there is no evidence for any variations in the shapes of the spectral lines in HD~102272 spectra. Neither BVS nor BC show correlation with radial velocity.
\vskip-0.15in

\section{Acknowledgements}
We acknowledge the financial support from the Polish Ministry of Science and Higher Education through grant 1P03D-007-30. GN is a recipient of a graduate stipend of the Chairman of the Polish Academy of Sciences. The Hobby-Eberly Telescope (HET) is a joint project of the University of Texas at Austin, the Pennsylvania State University, Stanford University, Ludwig-Maximilians-Universit\"at M\"unchen, and Georg-August-Universit\"at G\"ottingen. The HET is named in honor of its principal benefactors, William P. Hobby and Robert E. Eberly.


\end{document}